# Openness and Impact of Leading Scientific Countries

Caroline Wagner[1]*, Travis Whetsell[2], Jeroen Baas[3], Koen Jonkers[4]

**Abstract**

The rapid rise of international collaboration over the past three decades, demonstrated in coauthorship of scientific articles, raises the question of whether countries benefit from cooperative science and how this might be measured. We develop and compare measures to ask this question. For all source publications in 2013, we obtained from Elsevier national-level full and fractional paper counts as well as accompanying field-weighted citation counts. Then we collected information from Elsevier on the percent of all internationally coauthored papers for each country, as well as Organization for Economic Cooperation and Development (OECD) measures of the international mobility of the scientific workforce in 2013, and conducted a principle component analysis that produced an openness index. We added data from the OECD on government budget allocation on research and development (GBARD) for 2011 to tie in the public spending that contributed to the 2013 output. We found that openness among advanced science systems is strongly correlated with impact—the more internationally engaged a nation is in terms of coauthorships and researcher mobility, the higher the impact of scientific work. The results have important implications for policy making around investment, as well as the flows of students, researchers, and technical workers.

Keywords: research and development; national rankings; impact measures; openness; mobility

[1] John Glenn College of Public Affairs, The Ohio State University, Columbus, OH, USA 43210 (corresponding author)
[2] Steven J. Green School of International and Public Affairs, Florida International University, Miami, FL, USA
*Correspondence: Caroline Wagner, John Glenn College of Public Affairs, The Ohio State University, Columbus, Ohio, 43210, USA.
[3] Head of Data Science, Elsevier Research Intelligence, Amsterdam 1043 NX, NL
[4] Knowledge for Finance, Innovation and Growth, European Commission's Joint Research Centre, Brussels, BE

## 1. Introduction

For more than 50 years, organizations have measured science and technology (S&T) at the national level to assess and compare strengths (Godin 2002). The U.S. National Science Foundation in the 1950s and the multinational Organization for Economic Cooperation and Development (OECD) in the 1960s began collecting data on S&T, guided by the Frascati Manual (OECD 1963, 2015), to create indicators of activity. More recently, the United Nations Educational, Scientific and Cultural Organization (UNESCO) has issued decadal reports on world science (UNESCO 2010), reporting data for more countries than the OECD, but generally applying similar rules for data collection and reporting. These indicators and supporting statistics were developed for use by states, with the goal of accounting for public spending and informing future investment. Finnemore (1996) points out that international organizations such as OECD and UNESCO were established to serve states and do not address international interests.

In the decades since statistics and indicators were instituted, R&D has increasingly taken place across national boundaries, demonstrated by the rapidly increasing numbers of internationally coauthored articles (Adams 2013; Wagner et al., 2015). In unweighted terms, international coauthorships can account for as much as 60% of articles for some small countries. When normalized based upon fractional counts (where each country is given a proportional share of coauthored papers) these percentages drop to, on average, 25% of papers from the OECD countries. No official statistical measure accounts for international collaboration, nor do any economic measures exist of its contribution to growth, so any estimate of spending would be highly ambiguous.

This leaves a gap in scholarship and in policy because no clear accounting can be made of the increasing contribution of collaborative, team-based, and/or 'big science' projects. The amount of public funds going to international collaborations is mostly unknown. Budgetary measures for science have been made chiefly of inputs (Godin, 2002); most national budgets avoid line items for international S&T investments. (The European Union is the exception here because that organization can account for chiefly intra-European investments.) For some governments (notably, the United States), international investments are viewed as diffusing funds needed to build national capacity.

This article seeks to address this gap in assessment by suggesting a measure for the impact of international collaboration in science using fractionalized field-weighted citations and analyzing these in relationship to public spending and researcher mobility. We follow the recommendation of Moed and Halevi (2015) who suggest that in cases of complex accountability, a multidimensional approach should be used. Accordingly, we combined data sets (see Moed and Halevi, 2015, Table 2, p. 1994) using citation indexes and OECD national statistics to propose a measure of the benefits of international collaboration. We complement Taylor's (2016) approach to measurement – where he focused on broader economic links-- we limit our measure to publicly funded scientific ties, whether formed through international collaboration or through international mobility. This paper, apart from showing how interconnectedness is correlated to scientific impact, applies a novel way of applying the fractional counts of field-weighted citation to assign the impact of international collaborative research to the different partner countries as a way to assess impact.

This paper is an elaboration of a commentary published by Wagner and Jonkers in the journal Nature (2017). It attempts to stay close to the original analysis though it does include two additional European countries (Estonia and Slovenia). The fractional FWCI was compared in two ways: one by aggregating at the field level, and the second by using Elsevier's "all subjects" calculation. The field level calculation was conducted to account for citation differences among fields, which is considerable (Leydesdorff & Shin, 2013). This field level calculation was reported in *Nature* (Wagner & Jonkers 2017). However, this method led to over-counting of the number of articles because, in some cases, Elsevier assigns papers to more than one of 27 fields. In a similar vein, it led to higher values for the fracFWCI than would be the case with single counts. For this paper, we used the "all subjects" total that counts all articles just once, and then we compared the two measures. The resulting outcomes are similar for both methods in terms of the correlations and the explanatory power of the model.

   2. **Literature review**

Internationally coauthored articles account for close to 25 percent of Scopus articles, similar to the Web of Science (Wagner et al., 2015). Scholars have shown that internationally coauthored articles are more highly cited (Narin et al., 1991; Glänzel & DeLange, 2002; Ganzi et al., 2012), meaning these papers get more attention from the scientific community. The increased attention appears to be consistent across scientific fields (Wagner et al., 2015). Citations are considered a measure, not of quality, but of impact. To be consistently applied, impact has to be normalized in terms of fields (Leydesdorff & Shin, 2011), which is the process we used here. In this article, fracFWCI is a weighted average, where each article has a weight equal to the proportion of authors that represent the entity (country) on the paper. Elsevier provided these data from Scopus.

Appelt et al. (2015) studied the relationship between mobility, collaboration, and impact, showing that brain circulation is a "complex and multi-dimensional phenomenon…" but one that contributes to international coauthorship. Similarly, Sugimoto et al. (2017) and Franzoni et al., 2012, 2015, 2017 show that researchers who move from one country to another have a significant increase in citations to their work. Immigration and return flow of scientists contributes positively to scientific development in a given country (see also Baruffaldi & Landoni, 2012; Jonkers & Tijssen, 2008; Jonkers, 2010, Jonkers & Cruz-Castro, 2011, Jöns, 2009; Fernandez-Zubieta, Geuna & Lawson, 2015). The argument that outbound flows of scientists could also be positively associated with more impactful science production is perhaps a bit less obvious (in the context of "brain drain" and "brain gain" debates), but also appears to be a positive feature over time (Wagner 2009; Appelt et al., 2015).

Some literature suggests that the emigration of the highly skilled brainpower (including scientists) can be good for a country's economic development (Beine, Docquier and Rappoport, 2008; Papakonstantinou, 2017). Apart from remittances, strengthened ties and the potential positive effects of returnees, positive effects may also be due to the incentives provided by the potential to move, and the fact that successful emigres can be an inspiration for students. As Oded Stark and colleagues (1997) argued, potential migrants may invest more in the development of their own human capital if this investment makes it more likely for them to emigrate successfully. By analogy, scientists could have greater incentives to publish more and higher impact papers. They may invest in the development of their own scientific and technical human capital (Bozeman et al., 2001) if this is required to have access to mobility opportunities which helps in their scientific careers. Because expatriate scientists cooperate with their native countries at a high rate (Wagner, 2009), they also contribute to the development of their home system from abroad (Agrawal, Kapur, McHale & Oettle, 2011). This effect may have become stronger as a result of the developments in international information and communication infrastructures (Ding, Levin, Stephan and Winkler, 2010). Finally, as a developing country grows in capacity, emigrees may return to their system of origin. This movement of people, and its indirect contribution to science in a country of origin, also complicates efforts to account for benefits to the home country.

Assessment and evaluation methods are widely used by governments to ensure the efficiency and effectiveness of spending, but these are challenged by international collaborations. The flows of people between countries, and the connections across geographic distances, challenge the traditional methods of evaluation because it is difficult (if not impossible) to say where work is done or how credit should be assigned. As Moed and Halevi (2015) point out, stakes have grown for evaluation and assessment of research and development (R&D) as public budgets have become increasingly constrained. Yet, where international work is concerned, little progress has been made. Geographic distribution of researchers and research collaborations add complexity to any effort to account for performance of individual, institutions, or nations.

Since the early 1960s, economists have attempted to analyze the link between S&T investments and growth (Arrow, 1962; Schmookler, 1966; Freeman, 1982; Perez, 1983; Nelson, 1993). Taylor (2016) recently argued that differences in scientific and technological competitiveness of nations cannot be explained with reference to institutional differences alone, as might be assumed in a National Innovation Systems approach. Taylor said that the combination of the quality and intensity of domestic and international connections (networks) do explain a large degree of the difference between countries. This resonates with our own observations, and our research shows similar benefits.

To create a measure related to international collaboration, Wagner and Jonkers (2017) introduced an openness index which combines measures of scientific co-publication and mobility indicators. This article explains the openness index and provides greater details and documentation of this analysis to implement the Moed and Halevi (2015) recommendation to examine research input or capacity together with evolution of the number of active researchers in a country, combined with national statistics. Three measures are applied to the question of public spending and impact to test whether nations benefit from international collaboration: 1) the extent of international research collaboration by nation, 2) rates of researcher mobility, and 3) the scientific impact of national science as measured by field-weighted citation counts.

We acknowledge the complication of measuring the benefits to a nation of participating in international exchange and collaboration based upon bibliometric measures. National budgeting practices do not ease this task, since much of spending on international collaborative activities is buried in missions or grant-based funding in multiple programs and projects. Very few national budgets denote "international collaboration in science" as a line item. This is partly because international projects are not always listed in formal budget requests. These projects are likely to be challenged by a secular assembly, which face many competing priorities for funds. 'International' projects can be challenged and they are often removed from budget requests because there is no natural constituency for them.

Governments invest in international collaborative R&D in direct and indirect ways. Of course, some high-visibility, large-scale scientific organisations like CERN, ITER, and C-Band All Sky Survey receive direct government support for capital expenses. (CERN, for example, spends most of its budget on building and maintaining equipment, while only partially funding experiments.) In other cases, government ministries create funds to augment national R&D projects to include support for international cooperation; this is the case with the U.S. National Science Foundation's Office of International Science and Engineering (OISE[5]). OISE supplements grant-based R&D or dissertation research that can be enhanced by international collaboration. In still other cases, philanthropic organizations fund international cooperation, but this is a small percentage of the total.[6]

The exception to this model is the European Commission in the European Union and its EU level Framework Programmes through which it spent nearly 80 billion Euros over seven years – an amount which is likely to increase substantially in the next planning period. An important share of this Framework Programme funding is devoted to projects that require international collaboration between researchers in different EU Member States. The programmes are also open to third country, i.e. non EU Member States as the Commission´s research and innovation policy explicitly aims to be "open to the world."

Even in cases where international research projects involve applications for funding with well-defined and scrutinized research plans, clarifying exactly who is paying for what is challenging, and rarely done. In many countries, researchers maintain some discretionary power over resources, which they may devote to international activities. This is especially true in less resource-intensive fields of science where work is not tied to equipment or resources, but to a more simple exchange of ideas. The variety of approaches to research and to public budgeting complicate any efforts to tie funding to outcomes.

## 3. Data and Methods

This project compiles nation-level data on 35 nations from three different sources[7]. The analysis is limited to 35 nations because comparable data on mobility and government R&D spending (Government Budget Allocation in R&D-GBARD) were available only for these nations. The nations in this sample represent economically developed countries with strong systems of R&D. We recognize that this is a limited sample. However, we note that these nations represent about 92% of all publicly funded research. Follow-on research will add more nations to the list as data become available.

The main source of bibliometric data comes from Elsevier´s bibliographic database, Scopus publication and citation databases. We worked with Jeroen Baas, Senior Data Scientist at Elsevier, to derive a number of metrics from this database that require access to the full dataset for calculations. Other data were collected from the OECD Science, Technology, and Innovation Outlook 2016 and associated MSTI database[8] and national data sources.

### 3.1 Bibliometric data

Elsevier provided the bibliometric data for all articles indexed for 2013, with specific calculations of the fractional number of publications, where, in the case of internationally coauthored papers, each country with an address in the paper gets a proportional share of authorship. The fractional number of international papers was used to calculate the percentage of all papers that are internationally coauthored, by country. A second data set included the fractional Field-Weighted Citation Impact (FWCI) for each country, with citations for five years. The *fractional* Field-Weighted Citation Impact (FWCI) for a set of N publications belonging to entity $y$ is defined as:

$$fracFWCI \equiv \frac{\sum_{i=1}^{N}(\frac{c_i}{e_i}f_i)}{\sum_{i=1}^{N}f_i}$$

---

[5] As an example, NSF lists the international programs that are eligible for funds.
https://www.nsf.gov/od/oise/europe/sample_programs.jsp
[6] The Science Philanthropy Alliance conducted a survey in 2015 and estimated (based on a small sample) that philanthropic funding is about 5% of basic research funding in the United States (Williams, 2016).
[7] The scatterplot below includes Estonia and Slovenia for comparison purposes. These two nations were not included in Wagner & Jonkers, 2017
[8] The OECD database: Main Science and Technology Indicators, is at http://www.oecd.org/sti/msti.htm

$c_i$ = citations received by publication $i$ within a five year window
$e_i$ = expected number of citations received by publication $i$ within a five year window, based on all similar publications
$f_i$ = proportion of authors on publication $i$ belonging to entity $y$

The FWCI refers to "the ratio of citations received relative to the expected world average for the subject field, publication type and publication year" (SciVal, 2017). For example, a score of 1.50 means the publication receives 50 % more citations than the world average, a score of 0.50 means it receives 50 % less than the world average (Van Raan, 2005; Leydesdorff & Shin, 2011; Leydesdorff et al., 2013) when calculating values for individual articles, as it is an article-level metric. The FWCI values for countries are derived by aggregating the article-level values. In full counting, each article would count as one for each contributing country in publication counting, and the FWCI would be the average of the article level FWCI values. Elsevier aggregated the FWCI values for countries proportionally, generating a fractional FWCI value by assigning a weight to the article FWCI values according to the frequency with which a country appears in the authors' addresses on the paper. For example, an article with three countries contributing with each an equal number of authors on the paper, would weigh as 1/3rd in the calculation of the weighted average FWCI for each country. If the number of authors are different for each country, the weight is distributed proportionally. For instance, a paper with 2 authors from country A and 1 author from country B would assign a weight of 2/3[rd] to A and 1/3[rd] to B. In this article, fractional FWCI is a weighted average, where each article has a weight equal to the proportion of authors that represent the entity (country) on the paper.

The fractional FWCI was compared in two ways: one by aggregating at the field level, and the second by using Elsevier's "all subjects" calculation. The field level calculation was conducted to account for citation differences among fields, which is considerable (Leydesdorff & Shin, 2013). This field level calculation was reported in *Nature* (Wagner & Jonkers 2017). This method led to over-counting of number of articles because, in some cases, Elsevier assigns papers to more than one of 27 fields. In a similar vein, it led to higher values for the fracFWCI than would be the case with single counts. For this paper, we used the "all subjects" total that counts all articles just once, and then we compared the two measures. The resulting outcomes are very similar for both methods in terms of the correlations and the explanatory power of the model.

### 3.2 Mobility Scores and Openness Index

The OECD has collected data from each country about their research workforce, reporting number of new inflows, returnees, outflows, and percent of immobile workers. Elsevier presented the methodology in 2011 (see also Moed et al., 2013), to estimate different types of international mobility patterns on the basis of changes in scientific author affiliations in the period 2007-2013. We used data for 2013 on the percent of mobile researchers, new inflows, and returnees. OECD identified four different groups of mobility patterns. The "inflows" or immigrant scientists refers to the share of the authors which started publishing with an affiliation containing the country under study while initially using a different country as their institutional address. "Outflows," or emigrant scientists, refers to the share of researchers that started publishing with the country under study as their institutional address followed by publications indicating (an) other(s) country. "Returnees" refers to the share of the authors that first published from the country under study, followed by publications from a different country, and finally publishing again with the country under study in their institutional address. Mobile refers to all those researchers who have not remained in the same country during the observation period. The Mobility analysis by OECD using Scopus data is based the full career path since 1996 of all authors in Scopus with more than one publication. Using Web of Science data Sugimoto et al. (2017) operationalized international connection in a different (but related) way by analyzing individual mobility. Since the rest of this analysis is built on Scopus data, we decided to use the exploratory estimates on the share of these types of mobile researchers published by the OECD (2015). The data are available on figshare[9].

The openness index was developed using the mobility data and the percent of international coauthored articles based upon fractional counting. We inspected these data about mobility for each country to explore the flow of researchers as a factor related to openness. The indicators of mobility and engagement were found to be highly correlated with each other. As a result, we calculated a Principal Component (PCN) index of the four measures to create a single measure we called "openness" to indicate the extent of international engagement. The results of the PCN are shown in **Table 1** in the results section.

### 3.3 Government Spending

The project focuses on accountability for public spending, so we used OECD data on Government Budget Allocations or Outlays on R&D (GBARD) by country for 2011. The data are derived from OECD[10] and Eurostat, and in a few cases (e.g. China and Singapore) from national sources. GBARD is generally about 30 percent of total national spending (GERD). The justification for using GBARD is to limit the analysis to government spending. Government R&D spending is more likely than all spending to result in scientific publications. Use of GBARD reduces the chances that industrial spending on R&D would be counted, although there may be some number of articles that are funded by industrial R&D funds.

### 4. Results

The empirical approach taken in this research project is, first, to present bivariate correlations to analyze the relationships between impact, openness indicators, government financial support for R&D, and number of publications. Next, to economize the analysis, we use principal component analysis to combine several indicators of openness into single component variable for openness. Finally, we use linear regression to test the relationship between the openness component and impact, controlling for R&D funding and number of publications.

---

[9] https://figshare.com/articles/Spreadsheet_of_data_comparing_international_output/5082718
[10] http://www.oecd-ilibrary.org/science-and-technology/data/oecd-science-technology-and-r-d-statistics/government-budget-appropriations-or-outlays-for-rd_data-00194-en

First, **Table 1** shows the bivariate correlations between all the variables used in the analysis. The results show strong correlations between FracFWCI and the openness component, as well as each of the four indicators of openness; the strongest correlation is between international percent and inflows and mobile. The results show a very strong correlation between GBARD and fractional publication rate, but GBARD does not show a strong relationship with FracFWCI.

**Table 1 – Correlations**

Table 1

|  | 1 | 2 | 3 | 4 | 5 | 6 | 7 | 8 |
|---|---|---|---|---|---|---|---|---|
| 1 - FracFWCI |  |  |  |  |  |  |  |  |
| 2 - GBARD | 0.1137 |  |  |  |  |  |  |  |
|  | 0.5091 |  |  |  |  |  |  |  |
|  | 36 |  |  |  |  |  |  |  |
| 3 - FracPubs | 0.02679 | 0.84845 |  |  |  |  |  |  |
|  | 0.8767 | <.0001 |  |  |  |  |  |  |
|  | 36 | 36 |  |  |  |  |  |  |
| 4 - Int. Perc. | 0.76846 | -0.27492 | -0.36761 |  |  |  |  |  |
|  | <.0001 | 0.1046 | 0.0274 |  |  |  |  |  |
|  | 36 | 36 | 36 |  |  |  |  |  |
| 5 - NewInflows | 0.72562 | -0.10613 | -0.15425 | 0.78941 |  |  |  |  |
|  | <.0001 | 0.544 | 0.3763 | <.0001 |  |  |  |  |
|  | 35 | 35 | 35 | 35 |  |  |  |  |
| 6 - Returnees | 0.46826 | -0.21704 | -0.26163 | 0.68445 | 0.57691 |  |  |  |
|  | 0.0046 | 0.2104 | 0.129 | <.0001 | 0.0003 |  |  |  |
|  | 35 | 35 | 35 | 35 | 35 |  |  |  |
| 7 - Mobile | 0.73998 | -0.12949 | -0.19158 | 0.77385 | 0.97498 | 0.65189 |  |  |
|  | <.0001 | 0.4516 | 0.263 | <.0001 | <.0001 | <.0001 |  |  |
|  | 36 | 36 | 36 | 36 | 35 | 35 |  |  |
| 8 - Outflows | 0.69447 | -0.11399 | -0.17396 | 0.80007 | 0.94554 | 0.71213 | 0.97018 |  |
|  | <.0001 | 0.5144 | 0.3176 | <.0001 | <.0001 | <.0001 | <.0001 |  |
|  | 35 | 35 | 35 | 35 | 35 | 35 | 35 |  |
| 9 - Openness | 0.68197 | -0.26361 | -0.33819 | 0.85347 | 0.9335 | 0.80505 | 0.96064 | 0.957 |
|  | <.0001 | 0.126 | 0.0469 | <.0001 | <.0001 | <.0001 | <.0001 | <.0001 |
|  | 35 | 35 | 35 | 35 | 35 | 35 | 35 | 35 |

Top row = coefficient, middle row = p-value, bottom row = sample size

To economize the analysis, we used a principal component analysis approach to combine four indicators of national "openness" that were highly correlated into a single component variable. The openness component was included in the correlation **Table 1** to show the bivariate correlations. Principal component analysis is a common method to aggregate numerous theoretically related variables into single principal components (Dunteman, 1989; Nardo et al., 2005; O'Rourke & Hatcher, 2013). The eigenvalue for the principal component is 3.3 with the proportion of the variance accounted for by the component at 0.81. Other PCA models were tested, and no model showed a second component with eigenvalues above 1.0, so the number of components were set to one. The component loading in **Table 2** below show that each of the variables loads positively around 0.5 on openness. We took these results as an indication that the four variables were suitable for combination into a single principal component, and the factor scores were extracted for use alongside the other variables of interest.

**Table 2 – Component Loadings on Openness**

| Eigenvectors | Openness |
|---|---|
| International Perc. | 0.504435 |
| mobile | 0.531304 |
| Newinflows | 0.519326 |
| Returnees | 0.439957 |

Next, we show the relationship between openness and impact in a scatterplot to illustrate the relative positions of the nations. **Figure 1** shows three data points: 1) the X axis shows the openness index of a country based upon internationalization and mobility; 2) the Y axis shows the impact of a country's work by graphing the fractional FWCI of a country's publications; and 3) bubbles whose size is proportional to output (number of publications using fractional counting). The top right quadrant shows those countries that are both open and have a high fractional FWCI.

Notably, Switzerland, while small in geography and output, is high in both openness and impact. Singapore also appears very high in measures of openness and impact. These high performers are joined by the Netherlands, Denmark, Ireland, Belgium, and the United Kingdom in the high/high quadrant. Portugal also has a strong showing, perhaps reflecting policy changes to encourage greater R&D and engagement within Europe (Patricio, 2010).

Among the lowest performers in terms of openness and impact are China, Japan, and Turkey, as well as Russia. Surprisingly, South Korea is in the lower quadrant despite spending among the highest percentage of GDP on R&D. The USA has a positive position relative to impact, but a lower showing on openness, perhaps because of the large size of its scientific enterprise. Italy is less open than other European neighbors, but still shows relatively strong impact. Note, Slovenia and Estonia were added to the sample of 33 scientifically advanced countries used in Wagner & Jonkers (2017).

**Figure 1 – Scatterplot of FracFWCI and Openness**

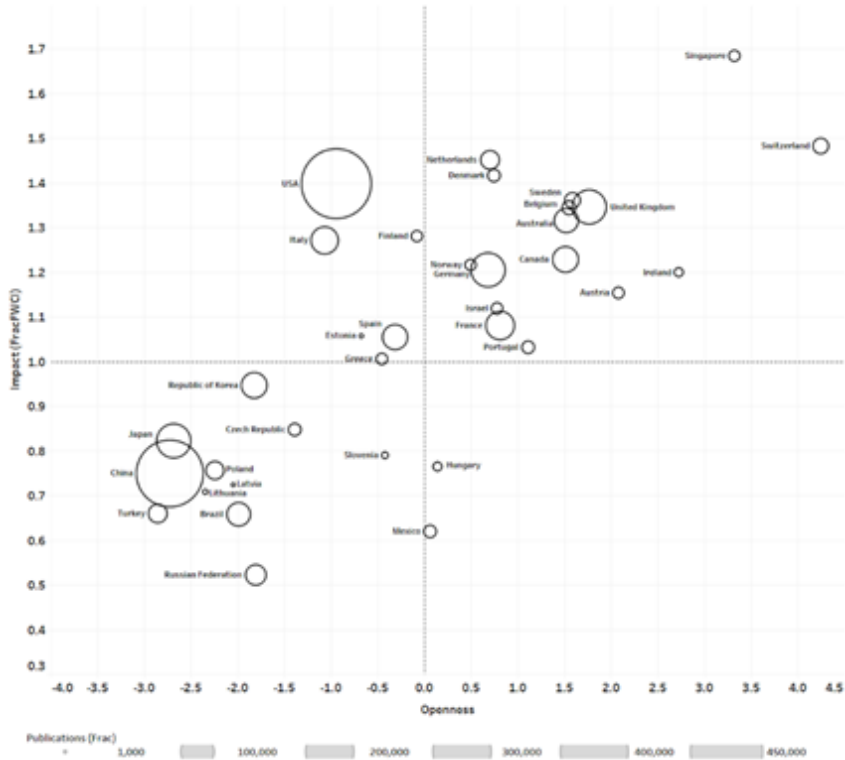

Finally, we conducted a linear regression analysis to account for control variables in the relationship between openness and impact in **Table 3**. We included two controls that are theoretically relevant to the impact of a nation's body of research. The first is the level of government funding for R&D (GBARD), and the second is the publication output of the nation (FracPubs). The parameter estimates are standardized to show the relative strength of the variables.

Openness shows a strong relationship with fracFWCI. GBARD and fracPubs are not significant with fracFWCI. The standard errors on these latter two variables however are inflated from collinearity, with a variance inflation (VIF) of around 3.5 and so should be taken with caution. We ran separate models with each variable separately, and they both displayed positive and significant correlations with fracFWCI individually with openness in the model. We decided to keep both variables in the model, because the model fit was better and because we are only using them as controls rather than key variables of interest (Allison, 2012). Standardized parameter estimates are shown in the first column, where openness has a value of 0.78 with impact. Government R&D shows a coefficient of 0.26. The proportion of variance, adjusted-$R^2$, in FracFWCI explained by the model is 0.53. These results provide stronger evidence of the positive relationship between openness and impact. Again, due to the small sample and the limited model, these results should be taken as preliminary.

**Table 3 – Linear Regression Analysis – Dependent Variable is Impact (FracFWCI)**

Table 3

| Variable | Stand. Est. | Parameter Est. | Std. Error | t Value | Pr > \|t\| |
|---|---|---|---|---|---|
| Intercept | 0 | 1.01373 | 0.04037 | 25.11 | <.0001 |
| Opennes | 0.77953 | 0.23107 | 0.03718 | 6.21 | <.0001 |
| GBARD | 0.26333 | 3.14E-06 | 2.65E-06 | 1.18 | 0.2453 |
| FracPubs | 0.08319 | 2.49E-07 | 6.83E-07 | 0.37 | 0.7176 |
| Adj. R-2 | 0.5273 | | | | |
| N | 35 | | | | |

### 5. Limitations

The results of the empirical analysis is preliminary and limited for several key reasons. First, the design is cross-sectional: panel data are not yet available on fracFWCI and mobility for the nations over time. This important caveat means that we are merely reporting cross-sectional correlations, rather than making claims about causality. As such, we can't rule out endogeneity between impact and openness, i.e. perhaps success breeds researcher mobility. Next, we included only a limited set of variables in the regression model. GBARD is considered by many experts to be the ideal variable for capturing government funded R&D. However, OECD data availability for GBARD is currently limited to about 30 nations. Notably, OECD's GBARD does not report a number for China or Singapore—the numbers used here were provided by national governments. Further, the mobility data are limited to a small set of developed nations. Thus, while we have bibliometric data (such as FWCI, number of pubs, and share of international co-publications) on a much larger set of nations, the analysis is limited by the measures of mobility and the data on government funding. Finally, we did observe collinearity between the fractionalized number of publications and GBARD, which inflates the variance of these variables. Thus, future research might seek to normalize GBARD by some other metric, or use a different measure altogether to disentangle collinearity with number of publications.

Using citation analysis also includes problems related to counts. To the question of citations as impact measures, we note an extensive and lively literature on this topic, reviewed in detail by Kostoff (1998). Among the limitations of citation analysis is the well known Matthew Effect (Merton 1968) that shows that citations go to those already cited; this affects all citation analysis. Furthermore, Katz and Martin (1997) argue that coauthorship is only a partial indicator of collaboration, since not all coauthorship represents cooperation Further, there is a phenomenon of multiple assignment of addresses, where authors list more than one address (Glänzel 2001), which may account for as much as 6% of coauthorship data. Hottenrott and Lawson (2017) found that the practice of listing multiple affiliations has doubled in the past few years (which further supports the idea that researchers are moving place more frequently). Finally, it is recognized that coauthorship counts are distorted by the occurrence of hyper-authored papers with more than 100 coauthors listed in the address lines (Cronin 2001; Kahn 2017). These problems are inherent in the data used. Although there are attempts (such as ORCID) to create unique identifiers for researchers, no norms exist for listing multiple affiliations in articles: some authors list addresses of institutions they are visiting for an extended stay; some list two addresses (home & visiting); some list just the home address, even when they are conducting research elsewhere. This means that some international links may be over counted if two addresses belong to one person, but it may also be undercounted if visiting scholars list only the address of a host or home institution.

Finally, we recognize that the numbers applied to countries are very highly aggregated, and that information has been lost as measures have been indexed and aggregated. We have been careful not to claim that the results are conclusive or demonstrate causation. We do believe that the results provide an interesting new approach to understanding the impact of international cooperation and we look forward to further testing of the approach and discussion about the findings.

## 6. Discussion and Conclusions

This paper expands on a recent commentary (Wagner and Jonkers, 2017) that introduced an openness index to provide input to measuring the benefits to nations of participating in international collaboration in science. The openness index is built upon measures of international collaboration evidenced by coauthorships and the several measures for the degree of the international mobility of a nation's scientists. These are then compared to five years of citations, fractionally allotted to nations. The goal of the work is to assign proportional shares of output and impact to nations, to link that to spending, and to use the analysis to gain insight into the impact of international collaboration and mobility (engagement) on the field normalized citation impact of national publication output.

The findings suggest that the countries that are open to international engagement tend to produce scientific articles that have a higher impact than those countries that are less open. We recognize that impact is not always the same as quality, but it is an indicator of engagement and recognition: people are paying attention to the work being produced across national boundaries. Countries that are highly 'open', in the sense that their researchers participate actively in international co-publication events, tend to produce higher impact research. We suggest that this indicates a national benefit from participating in international collaboration. This relationship is seen most notably in Figure 1 by the higher impact/high openness of demographically smaller nations which cluster in the top-right quadrant of the scatterplot. Singapore, the United Kingdom, the Netherlands, Switzerland, Sweden and Denmark all scored highly on the measures of openness and impact. It may be that, in order to conduct world-class research, smaller countries must cooperate—since funding research across the board is expensive. These countries, however, have relatively high degree of scientific resources (world leading research organisations, highly skilled scientific manpower and funding). Partially as a result, they are able to collaborate internationally and attract mobile scholars and in doing so further enhance their potential for high impact science production.

The observed correlation between openness and citation impact does not imply that all countries in principle have the same degree of potential to open their system to the outside world. Under-resourced research systems for example may struggle to both attract foreign research or returnees and engage in a high degree of international collaboration. This paper is restricted to a set of scientifically advanced nations. Even among those, there are degrees of potential/agency to open up. It should be clear that the paper does not make claim about countries in other parts of the world, which have not been analysed.

The observation that openness and engagement are linked to impact does more than simply confirm findings that show citation gains for international collaboration. It suggests that scientific mobility and connectivity may be factors in in encouraging higher impact work. The European Union has built its R&D funding programs on the premise that collaboration may raise impact, and it appears to have borne fruit. Further, the findings suggest that face-to-face cooperation remains a critical component of scientific advancement, a feature that has been discussed in the literature (see, e.g., Nardi & Whitaker, 2002; Wagner, 2009).

The finding of a relationship between openness and impact causes us to reflect upon recent anomalies in the shifting positions of countries in terms of scientific output and leadership. Those nations that are less 'open' appear to be lagging in terms of impact. Japan, in particular, has seen output and citation impacts remain flat since 2000 (Adams, 2012); Japan is also among the least internationalized of leading nations. The lack of international engagement may be dragging on Japan's performance. Writing in 2010, Adams et al. (2010) noted that Japan has a well-established research enterprise, and world-class universities: "…[so] it is puzzling to the observer that the average rate of citation to its research articles in the internationally influential journals … is significantly below…[other nations]." Lack of 'brain circulation' may be the answer to this puzzle.

In contrast, small and medium sized nations with enhanced global engagement have seen significant jumps in impact (Leydesdorff et al., 2014). Notable among these – and in addition to the well-known leaders of Switzerland, the Netherlands, Denmark, the UK, and Sweden – Singapore, Portugal, Belgium, and Austria stand out as countries that have increased their global reach and impact, with enhanced attention to their research. It appears that cross-boundary engagement and mobility have had positive effects in Europe, in particular. The location of large-scale intergovernmental laboratories and equipment may have an effect. Their presence in a country will most likely increase both openness and impact.

It is notable that of the countries appearing in the top right quadrant of the scatterplot– those with high impact and high openness–are also actively engaged within the European Research Area (ERA). Within the ERA, European governments have been implementing measures to improve the performance of domestic research systems, while promoting both international collaboration and mobility. Strengthening intra-European competition and collaboration are also central aims of the EU's Framework Programmes, which include instruments focusing on the strengthening of excellence, intra-European mobility and the establishment of pan-European research consortia. Its Framework Programmes are open to participation from researchers based in non-EU countries - one of the current stated objectives of EU research policy is to be more "open to the world". The USA holds several anomalous positions at the global level. First, it has been noted that the United States has seen drops in percentage shares among highly cited publications. In fact, despite the EU overtaking the US in top 10% most highly cited publications, recent analyses by Leydesdorff et al. (2014), Rodriguez-Navarro & Narin (2017) show that the USA still leads in producing the top 1% most influential advances in science. This is the case even though, in percentage terms, the USA is less 'open' than other leading nations. The USA continues to attract scientists from around the world, but appears not to be sending a proportional number abroad. This may be because, compared to the other countries analysed, the United States is both huge and has a large home-grown scientific workforce that collaborates and moves freely and frequently between its constituents states (Economist, 2017). The size of the US system in combination with a home bias to citations (Frenken, 2009; Börner et al., 2006) might also result in inflated impact figures in comparison to smaller systems.

The correlation between openness and citation impact is strong, when controlled for by R&D funding and numbers of articles published. Countries with low openness and low impact include Russia, Turkey and Poland, China, Japan, Latvia, Lithuania, the Czech Republic and, against expectations, South Korea (which spends a higher percentage of its GDP on R&D than almost every OECD country, including the United States). These countries are shown in the lower-left quadrant. Mexico performs well below what might be expected from the observed correlation between openness and impact observed in the other countries. While an OECD country, Taylor (2016) argues that the lack of stable and sustained investment in its science system has reduced the effectiveness of national spending. Why Hungary and Italy perform differently than the other countries in our sample requires further study.

Policy actions to nationalize research practices and reduce international engagement would appear to be antithetical to impact, and perhaps to creativity. While we cannot draw a causal relationship between openness and impact based upon this analysis, the initial indication is 'brain circulation' may be critical to science by bringing fresh ideas, enhancing creativity, and raising quality, as suggested by Saxenian (2005) for India and China, by Jonkers & Tijssen (2008) for China, by Jonkers & Cruz-Castro (2013) for Argentina, by Baruffaldi & Landoni (2012) for Italy and by Jöns (2009) for post-war Germany.

Taylor (2016) argued that the scientific and technological prowess of nations is strongly related to their integration in international commercial networks. In a follow-on paper we will explore the relationship between scientific and economic openness and the extent to which various measures of economic, social, cultural or political openness can help generate a model that explains more of the variations in the performance of national research systems.

**7. Acknowledgements**

We thank Loet Leydesdorff and two anonymous reviewers for providing comments on an earlier draft. We thank Monya Baker at *Nature* Magazine for helping to create the commentary appearing on October 5, 2017. In addition, these findings were presented at the American Association for the Advancement of Science (AAAS) annual conference, February 2017 and we gained a good deal of benefit from comments by fellow panelists and the audience.